\documentclass{aa} 
\usepackage{xcolor}
\usepackage[hidelinks]{hyperref}
\usepackage[multiple]{footmisc}

\usepackage[normalem]{ulem}
\usepackage[version=4]{mhchem}

\usepackage{graphicx}
\usepackage{txfonts}
\defcitealias{vanlooveren2024}{VL24}

\begin{document}

   \title{Habitable Zone and Atmosphere Retention Distance (HaZARD)}
   \subtitle{Stellar-evolution-dependent loss models of secondary atmospheres}
   \titlerunning{HaZARD}

   \author{Gwena\"el Van Looveren \inst{1}
          \and
          Sudeshna Boro Saikia \inst{1}
          \and
          Oliver Herbort \inst{1}
          \and
          Simon Schleich \inst{1}
          \and
          Manuel G\"udel \inst{1}
          \and
          Colin Johnstone \inst{2}
          \and
          Kristina Kislyakova \inst{1}
          }

   \institute{Department of Astrophysics, University of Vienna,
                T\"urkenschanzstra{\ss}e 17, A-1180 Vienna, Austria\\
                \email{gwenael.van.looveren@univie.ac.at}
            \and
                International Institute for Applied Systems Analysis (IIASA), 
                Schlo{\ss}platz 1, A-2361 Laxenburg, Austria
             }

   \date{Received November 15, 2024; accepted January 31, 2025}

 
  \abstract
   {Thanks to the \textit{James Webb} Space Telescope (JWST), observations of the secondary atmospheres of rocky planets have become possible. Of particular interest are rocky planets orbiting low-mass stars within the habitable zone (HZ). However, no thick secondary atmospheres have been found around Earth-sized planets to date. This leaves open the question of whether secondary atmospheres are rare around Earth-sized rocky exoplanets.\ }
   {In this work we determine the distance at which an Earth-sized planet orbiting a variety of stellar hosts could retain a CO$_2$- or N$_2$-dominated atmosphere and compare this atmospheric retention distance (ARD) with that of the liquid-water HZ.}
   {We combined planetary atmosphere models with stellar evolution models. The atmospheric models produced by the thermo-chemical Kompot code allowed us to calculate the Jeans escape rates for different stellar masses, rotation rates, and ages. These loss rates allowed us to determine the closest distance a planet is likely to retain a CO$_2$- or N$_2$-dominated atmosphere. Using stellar rotation evolution models, we modelled how these retention distances evolve as the X-ray and ultraviolet activity of the star evolves.}
   {We find that the overlap of the HZ and the ARD occurs earlier around slowly rotating stars. Additionally, we find that HZ planets orbiting stars with masses under 0.4 M$_\odot$ are unlikely to retain any atmosphere, due to the lower spin-down rate of these fully convective stars. We also show that the initial rotation rate of the star can impact the likelihood of a planet retaining an atmosphere, as an initially fast-rotating star maintains high levels of short-wavelength irradiance for much longer.}
   {The orbits of all Earth-like rocky exoplanets observed by JWST in cycles 1 and 2, including HZ planets, fall outside the ARD. Our results will have implications for future target selections of small exoplanet observing programmes with JWST or future instruments such as the Ariel space mission.}


   \maketitle
%

\section{Introduction}

One of the key priorities of exoplanet science in this century is the discovery and characterisation of secondary atmospheres around rocky exoplanets\footnote{\href{https://www.nationalacademies.org/our-work/decadal-survey-on-astronomy-and-astrophysics-2020-astro2020}{ Astro2020 Decadal survey}}$^,$\footnote{\href{https://www.cosmos.esa.int/web/voyage-2050}{ ESA Cosmic Vision 2050}}. Of particular interest are rocky exoplanets in the habitable zone (HZ) of their host stars, a circumstellar distance at which a planet can sustain surface liquid water \citep{hart1979,kasting1993,kopparapu2013,kopparapu2014,ramirez2014}. In our Solar System, the HZ encompasses three rocky planets, Earth, Venus, and Mars. While present-day Earth has an N$_2$-dominated secondary atmosphere with surface liquid water, present-day Venus and Mars have CO$_2$-dominated atmospheres \citep{prinn1987}. Hence, the search for an Earth-like or a Venus-like atmosphere is focused on HZ rocky exoplanets around low-mass main-sequence stars (spectral type F to M). A prime example is the TRAPPIST-1 system in which at least three of its seven rocky planets lie in the HZ of its low-mass host star of spectral type M8 V \citep{gillon2017}.

A vast majority of observations have been aimed at rocky exoplanets around M-type stars, primarily using {\it Spitzer} and the {\it Hubble} Space Telescope \citep{wordsworth2022}, as these stars are ubiquitous and the planet-to-star radius ratio is more favourable for observations due to the small stellar radii. Despite this advantage, however, rocky exoplanet observations with these instruments led to non-detections or inconclusive results \citep{dewit2016,diamond-lowe2018,edwards2021,mugnai2021}. Considering that quite a few of these rocky exoplanets are located at short orbital distances from their host star, complete atmospheric escape is considered to be a primary cause for the reported non-detections, such as for LHS 3844 b \citep{kreidberg2019}. However, instrumental limitations or masking of an atmospheric signal due to cloud coverage or stellar contamination could not be ruled out.

The {\it James Webb} Space Telescope \citep[JWST;][]{gardner2006,gardner2023}, with its high sensitivity and near- to mid-infrared wavelength coverage, heralds a new era for studies of rocky exoplanet atmospheres. Thermal emission of rocky exoplanets observed by JWST's Mid-Infrared Instrument (MIRI) is expected to determine the presence or absence of a secondary atmosphere in $\sim$100 targets \citep{koll2019}. Additionally, transmission spectroscopy using JWST's near- and mid-infrared instruments is expected to identify molecules such as CO$_2$ and H$_2$O for the first time in such exoplanets, for example the TRAPPIST-1 HZ planets \citep{barstow2016,morley2017,krissansen-totton2018,lustig-yaeger2019}. 

 Thermal emission measurements of rocky exoplanets with JWST have so far led to atmospheric non-detections or inconclusive results. JWST/MIRI observations of TRAPPIST-1 c rule out the presence of a thick CO$_2$ atmosphere \citep{zieba2023}. Similar observations of TRAPPIST-1 b also suggest an airless rocky body \citep{greene2023}. \citet{ducrot2024} propose an alternative scenario for TRAPPIST-1 b that involves a hazy CO$_2$ atmosphere with temperature inversions. However, their analysis also hints at an airless body as an equally likely scenario for TRAPPIST-1b. Hence, simulations of the secondary atmosphere are essential to breaking the degeneracy of JWST/MIRI thermal emission observations.

In the search for a secondary atmosphere around rocky exoplanets, the HZ acts as a key requirement, as it indicates the potential presence of surface liquid water. Typically, HZ calculations are carried out using climate models \citep[e.g.][and we provide more details in Sect. \ref{sec:hz}]{kopparapu2013}. By definition, a HZ rocky planet is in principle capable of sustaining surface liquid water, provided it has an atmosphere and the right pressure-temperature profile. However, a location in the HZ of a star does not guarantee an atmosphere for a rocky planet observed today even if it had an atmosphere in the past. This is because the HZ calculations do not take into account that the magnetic activity of a star, in the form of high-energy X-ray and ultraviolet (XUV) radiation and stellar winds, can erode a planet's atmosphere, even in a largely varying manner over evolutionary timescales \citep{lammer2003,grasser2023}. 

Atmospheric loss driven by the star could be in the form of (a) thermal loss driven by XUV radiation, which usually comes in the form of Jeans escape or hydrodynamic escape, or (b) non-thermal atmospheric escape mainly due to stellar winds. In this work, we focus on mass loss through Jeans escape, which occurs when the velocity of a particle at the top of the atmosphere is greater than the escape velocity. Because the calculation of this loss mechanism assumes a Maxwellian velocity distribution, which implicitly assumes a hydrostatic atmosphere, we believe it represents a lower limit to the total mass-loss rate at any given time. Non-thermal loss processes can only add to the loss rate; for example, on Earth, Jeans escape is estimated to only make up 10 to 40\% of the total mass loss \citep{catling2017}. The atmosphere can become hydrodynamic for planets experiencing higher irradiation from their host star, making thermal losses in the form of hydrodynamic escape the dominant mass-loss mechanism \citep{tian2008a}. Studies of hydrogen-dominated atmospheres \citep[e.g.][]{erkaev2013,kubyshkina2018a} have shown that even for more massive planets, hydrodynamic mass loss can reach the so-called blow-off state, where entire atmospheres can be lost in less than a million years. 
As such, we assume that the other loss mechanisms would only increase the mass-loss rates beyond the calculated values.

Previous studies involving primordial hydrogen and helium atmospheres show that an entire atmosphere could be lost due to stellar XUV alone \citep{lammer2003,owen2012,owen2016}. When it comes to rocky planets' secondary atmospheres, heavier species such as CO$_2$, N$_2$, and H$_2$O are usually of interest. These heavier species also bring forth their own complexity through their chemistry, radiative transfer, and physical state (e.g. the presence of clouds and precipitation). This complexity is illustrated through CO$_2$, which has a warming effect on the lower atmosphere and surface. As a strong and common greenhouse gas, it is often used to define HZ boundaries \citep[e.g.][]{kasting1993}. On the other hand, the low exobase temperature of Venus compared to that of Earth, despite the former's higher insolation, demonstrates the cooling effect of CO$_2$ in the upper atmosphere \citep[e.g.][]{dickinson1972}.

Escape calculations for secondary atmospheres, for example CO$_2$, were first applied to Solar System rocky planets \citep{fox1991,roble1995}. The diversity in the known rocky exoplanetary population requires that such models be expanded and applied to different stellar and planetary conditions \citep{tian2008a,johnstone2018,nakayama2022}. By applying the model described in \citet{johnstone2018}, we recently calculated the atmospheric loss of CO$_2$- and N$_2$-dominated atmospheres around the TRAPPIST-1 planets \citep[hereafter VL24]{vanlooveren2024}. Our results show that the present-day XUV radiation of TRAPPIST-1 is sufficient to erode CO$_2$- and N$_2$-dominated atmospheres on all seven rocky planets within a few hundred million years. The calculations only considered Jeans escape and did not take the star's early evolution into account, which would have subjected the planets to even stronger XUV irradiation; the results thus provide a conservative approach to the mass loss over time. Nonetheless, these results suggest that, at present, none of the TRAPPIST-1 planets is suitable for surface habitability due to the lack of an atmosphere unless a secondary atmosphere is rapidly replenished through, for example, excessive volcanic activity \citep{luger2017, kislyakova2017}.

In \citetalias{vanlooveren2024} we calculated the minimum orbital distance from TRAPPIST-1 where a CO$_2$- or N$_2$-dominated atmosphere around a rocky planet would survive over hundreds of millions of years. We call this distance the `atmospheric retention distance' (ARD). TRAPPIST-1 is a very low-mass M star (spectral type M8 V), and as a result we cannot generalise these results to other exoplanetary systems, for example planets around F, G, K, and early M stars. This paper aims to extend this work towards Earth-sized rocky exoplanets orbiting low-mass main-sequence stars (from spectral types FV to MV) and study where the HZ and ARD overlap, or if HZ conditions and atmospheric retention can be mutually exclusive.

\section{Method}\label{sec:2}
In this section we present the various components required to build up to the ARD. First, we take a closer look at the literature on outgassing to determine the rate at which secondary atmospheres build up and get replenished. In Sect \ref{sec:planet} we discuss the atmospheric loss model, from which we can determine the irradiance when an atmosphere is lost. Next, we look at the luminosity of various stars to determine the distance where this critical loss irradiance occurs. To better understand these distances in the context of the planetary system, we discuss the calculation of the HZ in Sect. \ref{sec:hz}. Finally, we finish this section by summarising the parameter space that we cover in this work.

\subsection{Outgassing} \label{sec:outgassing}
First, we take a closer look at the formation and replenishment of secondary atmospheres. Not all volatiles of a planet will be in the atmosphere throughout its lifetime, but rather some of them will be retained in the mantle and outgassed over time \citep[summarised in Fig. 1 of][]{lichtenberg2023}. During the early planetary life outgassing through a global magma ocean, and in particular, its solidification, leads to the first buildup of a secondary atmosphere of tens up to thousands of bars \citep{elkins-tanton2008}. The depth of the magma ocean and the composition of the planet not only determine the thickness of the atmosphere but also its composition \citep[as summarised in][]{lammer2018}.

Rocky planets can outgas enough gas to build up a new atmosphere over longer timescales if the atmosphere has been lost previously due to the large mass loss. Alternatively, continuous outgassing is balanced by continuous loss over long timescales. For example, current outgassing rates of Earth associated with volcanism provide $1.0 \times 10^{4}$~kg/s (7.5 Tmol/yr) of CO$_2$ and $5.4 \times 10^{4}$~kg/s (95 Tmol/yr) of H$_2$O \citep[][p. 203]{catling2017}. As summarised in Sect. 5 of \citet{lammer2018}, the delivery of nitrogen to the atmosphere can also occur through volcanism in small amounts (less than 1\% of all outgassed material). For present-day Earth, the nitrogen cycle is inextricably tied to the biosphere.

Within the Solar System, the Earth seems to be an outlier concerning plate tectonics and the resulting geological activity. However, early Earth is believed to not have had plate tectonics. \citet{Guimond2021} show that for stagnant lid planets similar to the early Earth, the outgassing rates can reach and stay at about $1.4 \times 10^{4}$~kg/s ($10\,$Tmol/yr) and consist of mainly CO$_2$ with additional CO and H$_2$O. In another study, \citet{Baumeister2023} investigated the planetary interior and outgassing evolution of a large range of planets with different mantle redox states, volatile contents, masses, and sizes. In general, volcanic outgassing rates can stay at about $10^2$\,kg/s ($7\cdot10^{-2}\,$Tmol/yr) of CO$_2$ for multiple gigayears. The evolution model also allows us to investigate the timescale over which the rocky planets stay volcanically active, which is defined by the presence of melt in the mantle. Although these timescales are usually defined by the thermal budget of the planet at formation, other effects such as tidal and possibly induction heating can continue to influence the amount of melt \citep{kislyakova2017}.

To compare the outgassing rates, typically given in units of Tmol/yr, to the mass-loss rate at the ARD, the mean molecular weight of the outgassed material needs to be addressed. For the outgassing of pure CO$_2$, an outgassing rate of $1.7 \times 10^{4}$~kg/s ($12\,$Tmol/yr) equates to 0.1 $M_{\mathrm{atm,\oplus}}$ over 1\,Myr, where $M_{\mathrm{atm,\oplus}}$ is the mass of Earth's atmosphere ($5.14 \times 10^{18}$~kg, \citealt{Trenberth2005}). Once the overall outgassing rate is of the order of the mass-loss rate, a steady-state atmosphere can be retained. For outgassing rates higher than the loss rate, a new secondary atmosphere can build up. Throughout this work, we used this outgassing rate as an upper limit of outgassing rates that can be sustained over longer periods of time. However, the exact outgassing rate would depend on the specific properties of the planet.

Overall, we can conclude that outgassing from volcanism can sustain on rocky planets for billions of years before the planet becomes geologically dead, that is, when there is no longer melt in the interior to drive geological processes. We did not consider any chemical weathering of the surface rocks. For an old stagnant lid planet, this assumption equates to a fully weathered surface.

\subsection{Mass-loss model}\label{sec:planet}
In this study we used the Kompot model, which is a 1D thermo-chemical physical model for planetary upper atmospheres \citep{johnstone2018}. This code models the vertical structure of the upper atmospheres of planets. The modelled region reaches from 1~mbar, the lowest pressure where clouds typically form, up to the exobase, the limit above which the gas is collisionless. The code takes into account heating from the host star's XUV spectrum, cooling due to IR emission, and thermal conduction to determine the thermal profile. These changes are also influenced by changes in the chemical profile, which are modelled through both chemical and photochemical reactions, molecular and eddy diffusion, and advection. Using this model, we evolved the atmosphere using a time-marching scheme until it came to a steady state, which represents our solution. The important boundary conditions that determine this solution are the chemical composition and temperature at the lower boundary and the input stellar XUV spectrum at the upper boundary.

In \citetalias{vanlooveren2024}, we simulated planets with 0.8, 1.0, and 1.2 M$_\oplus$ to cover five out of the seven planets in the TRAPPIST-1 system. In the present work, we focused only on 1.0~M$_\oplus$ planets to study the relation between the ARD and the HZ without adding the additional complexity of planetary mass. We varied the atmospheric compositions from N$_2$-dominated (90\%~N$_2$, 10\%~CO$_2$) to CO$_2$-dominated (1\%~N$_2$, 99\%~CO$_2$) to represent common atmospheric compositions found around the rocky bodies in the Solar System and possible levels of cooling achieved by CO$_2$. These compositions limit the chemical network to species composed of C, N, and O.

Using the exobase altitude, temperature, and composition of these models, we can determine the Jeans mass-loss rate for each species \citep[e.g.][]{bauer2004}. These individual mass-loss rates are then averaged over the mixing ratios at the exobase. It is important to note that temperature at the exobase is not necessarily uniform over the entire planet. In fast-rotating atmospheres, such as that of Earth, the day and night temperatures at the exobase vary between ~1200~K and ~1100~K \citep{emmert2021}. On planets with a slowly rotating atmosphere, such as Venus, measurements have shown that the day-side upper atmosphere can reach temperatures around 300~K, whereas the night-side temperatures only go up to 150~K \citep{limaye2017}. For Earth, it was shown in \citet{johnstone2018} that assuming a 66$^{\circ}$ angle to the zenith is representative of the average thermal profile. To not overestimate the mass-loss rate based on a 1D simulation, we integrated the loss rates only over the day side of the planet and omitted additional, perhaps lower, loss rates from the night side.

\begin{figure}
    \centering
    \includegraphics[width=\linewidth ]{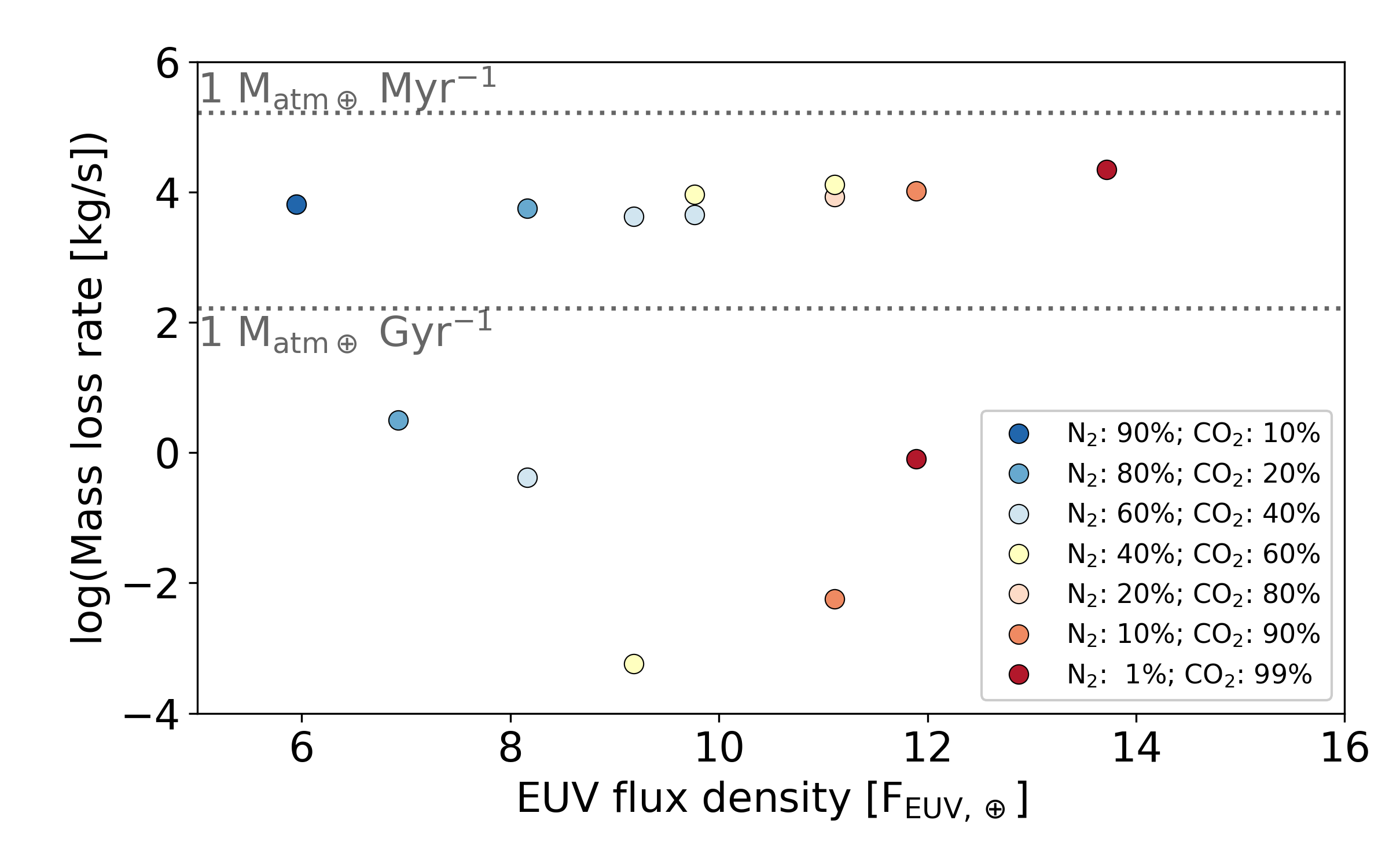}
        \caption{Abundance-weighted average Jeans escape for various atmospheric compositions, distinguished by colour \citepalias[extension of the results presented in][]{vanlooveren2024}.}
        \label{fig:mass_losses}
    \end{figure}

For each modelled atmospheric composition, we determined the irradiance at which Jeans escape becomes catastrophically large. In this study we defined this limit as when the entirety of Earth's atmosphere is lost within 10 Myr, or approximately $1.6 \times 10^{4}$~kg/s, which is of the same order of magnitude as some of the higher outgassing rates. This loss rate would be equivalent to a loss of 0.004 Earth's ocean in 10 Myr, assuming the mass of the Earth's ocean to be $1.4 \times 10^{21}$~kg \citep{wolf2015}. The mass-loss rates of these models are summarised in Fig.~\ref{fig:mass_losses}. For each type of atmosphere, we determined the extreme-ultraviolet (EUV) irradiance at which this catastrophic mass-loss rate is reached. These fluxes can then be used to determine the distance where such losses occur around different stars (i.e. the ARD).

It is important to note that at these irradiances the atmosphere is likely to be hydrodynamic. \citet{tian2008a} estimate an Earth-like atmosphere to become hydrodynamic at 5 times the current solar EUV flux ($\sim$25 mW/m$^2$). An atmosphere experiencing hydrodynamic outflow could reach losses much greater than those resulting from Jeans escape calculations. Due to this change in the dominant effect, we do not extrapolate our results beyond the simulated models. This further emphasises that these estimates would represent a lower limit of the real mass loss of a planet.

\subsection{Stellar spectrum}\label{sec:star}
The second important aspect to determine the ARD is the influence of the star. The modelled atmospheres were all exposed to an empirical model of the solar spectrum based on \citet{claire2012}. To model the thermal profiles we used the spectrum between 1 and 400 nm, as the photochemical reactions included in the Kompot code are particularly sensitive to this wavelength regime. Unfortunately, the interstellar extinction makes observations of stellar spectra between about 10 and 20~nm and 121~nm nearly impossible. Several studies \citep[e.g.][]{fontenla2016, namekata2023} have attempted to reconstruct the spectra of different stars in this wavelength range. However, the small number of stars with observable EUV spectra leads to great uncertainty of the strength of lines. As shown in Fig. 7 of \citetalias{vanlooveren2024}, the exact spectrum of a star can influence the pressure-temperature profile of a planet. Modelling all possible combinations of planetary atmosphere and stellar spectrum is outside the scope of this study and we instead scaled the solar spectrum uniformly to get spectra of stars with different activity levels. Since we aim to determine general trends in this work rather than specific cases, we believe this to be a reasonable approximation to make. Since most known planets orbit relatively evolved main-sequence stars, the spectral shape of an XUV spectrum is indeed supposed to be similar to that of the Sun \citep{johnstone2015a}.

To encompass a wide array of different stars, we used the stellar parameters from \citet{johnstone2021a}. In their work, the authors determined the evolution of the luminosity of stars in different wavelengths for various stellar masses and initial rotation rates. The wavelength range of importance for the current work is the EUV domain, which they define as between 10 and 92\,nm. It is important to note that in our previous work \citepalias{vanlooveren2024} we used a slightly different definition of the EUV domain, which was defined within the range 10-121\,nm. To relate the modelled atmospheres to the stellar models, we calculated the flux density of the solar spectrum between 10 and 92\,nm and redefined the EUV flux density at 1 au as $\mathrm{F_{EUV,\oplus}}= 1.1 \times 10^{28}\, \mathrm{erg/cm^{2}}$.

\citet{johnstone2021a} provide data for a large number of initial rotation rates, as this parameter strongly influences the XUV evolution of a star. In this work, we adopted the same approach to the initial rotation rates as \citet{johnstone2021a}, defining slow, medium, and fast rotators as the 5th, 50th, and 95th percentile, respectively, of the distribution of initial rotation rates. This way we aim to demonstrate the most likely range of possible scenarios without overcrowding graphs.

\subsection{Habitable zone}\label{sec:hz}
In this work we calculated the HZ according to the methods outlined in \citet{kopparapu2013} and expanded upon in \citet{kopparapu2014}. The latter work demonstrates the dependence of the HZ on planet mass, showing how more massive planets can lead to a wider HZ. Because we focused on Earth-mass planets, we only used the updated values provided for 1 M$_\oplus$ in the latter work.

In \citet{kopparapu2014} the inner edge of the HZ is determined by runaway greenhouse, a scenario in which oceans would evaporate entirely. The outer edge of the HZ is defined as the maximum greenhouse, the furthest point where CO$_2$ can keep the surface temperature above 273 K. Due to this definition, the boundaries can also vary on the atmospheric composition and particularly the CO$_2$ content. The authors emphasise that these limits are conservative and that the true HZ is probably wider. Throughout the rest of this work, we use these boundaries to define the HZ. This estimate of the HZ provides an upper limit estimate for the potential extent of the HZ under ideal circumstances. As we also varied the composition of the atmosphere to determine the ARD, it is important to note that this can lead to cases where a certain atmospheric composition could be retained in the HZ but not meet the conditions required for liquid surface water. Other effects, such as the 3D nature of planets or tidal locking, can further influence these boundaries, generally moving the inner boundary closer in \citep{kopparapu2016,chen2019}. As such, the HZ shown here functions more as a guide for scale in different stellar systems than an absolute boundary for habitability.

The equations to determine the HZ \citep[Eqs. 2 and 3 in][]{kopparapu2013} require the bolometric luminosity and effective temperature of the host star. These values for different stellar masses and ages were calculated following \citet{johnstone2021a}.

\subsection{Parameter space}\label{sec:ip}
In Table \ref{tab:grid} we summarise the parameter ranges that are covered in this study. This table serves to illustrate that we focused on Earth-like planets both in mass and atmospheric composition. 

The parameter range described in Table \ref{tab:grid} covers a large variety of stars and planets that are of interest in the search for habitable worlds. The mass range of habitable rocky worlds has been estimated to be as small as 0.0268 M$_\oplus$ \citep{arnscheidt2019} or as large as $\sim$6 M$_\oplus$ (1.6 R$_\oplus$) as the transition mass between rocky planets and sub-Neptunes \citep{rogers2015}. In this work we focused on planets with a mass of 1 M$_\oplus$. The stellar masses cover stars with masses of about the Sun's and lower, as heavier stars are much brighter and shorter-lived, making them poor hosts for habitable worlds. From an observational perspective, smaller stars are also preferable due to the smaller difference in radius of the planet compared to the star. The latest selection of targets for the Ariel mission, which will characterise exoplanet atmospheres, consists of stars between 0.09 and 2.77 M$_\odot$, though the majority of targets fall in the narrower range of about 0.6 to 1.5 M$_\odot$ \citep{edwards2022}. By studying this parameter range we aim to identify the planet--star combinations that are most likely to ensure the planets keep their atmospheres so that we can optimise the probability of finding Earth-sized planets with atmospheres and perhaps even biomarkers.

On top of covering various stellar masses, we also looked at a wide array of different stellar ages. The luminosities, both bolometric and EUV, can vary strongly over the lifetime of a star, which can lead a planet to experience different environments. This in turn can lead to a more constrained set of possible atmospheric parameters. \citet{johnstone2021} demonstrated how the Archean Earth required a minimum mixing ratio of 40\% CO$_2$ to maintain an atmosphere in the environment of the young Sun, which we discuss in Sect. \ref{sec:res_sol}. 

Lastly, in this work we focused on varying compositions N$_2$ and CO$_2$ as they are the main constituents of the atmospheres of rocky planets in the Solar System. It is important to note, as described in \citetalias{vanlooveren2024}, that these models do not include water. It was shown by \citet{johnstone2020} that water vapour easily results in large and hydrodynamic losses. Because of hydrogen's low mass, it is lost more easily than any other element and we believe it is justified to not include water as it would only increase the calculated mass-loss rates.

\begin{table}
\caption{Parameters covered by the input models.}
\label{tab:grid}
\centering
\begin{tabular*}{\linewidth}{l l l}
\hline\hline
Parameter        & Covered values   & Unit                      \\
\hline
Planet mass      & 1.0              & M$_\oplus$                \\
Atmospheric composition & 10/90 - 99/1     & \%~CO$_2$/\%~N$_2$          \\
Irradiance       & 6 - 14           & F$_{{\rm EUV,}\oplus}$    \\
Stellar mass     & 0.1 - 1.2        & M$_\odot$                 \\
Stellar age      & 1.0 - 12000      & Myr                       \\
\hline 
\end{tabular*}
\end{table}


\section{Results and discussions}\label{sec:res}

\subsection{Solar System} \label{sec:res_sol}
\begin{figure}
    \centering
    \includegraphics[width=\linewidth]{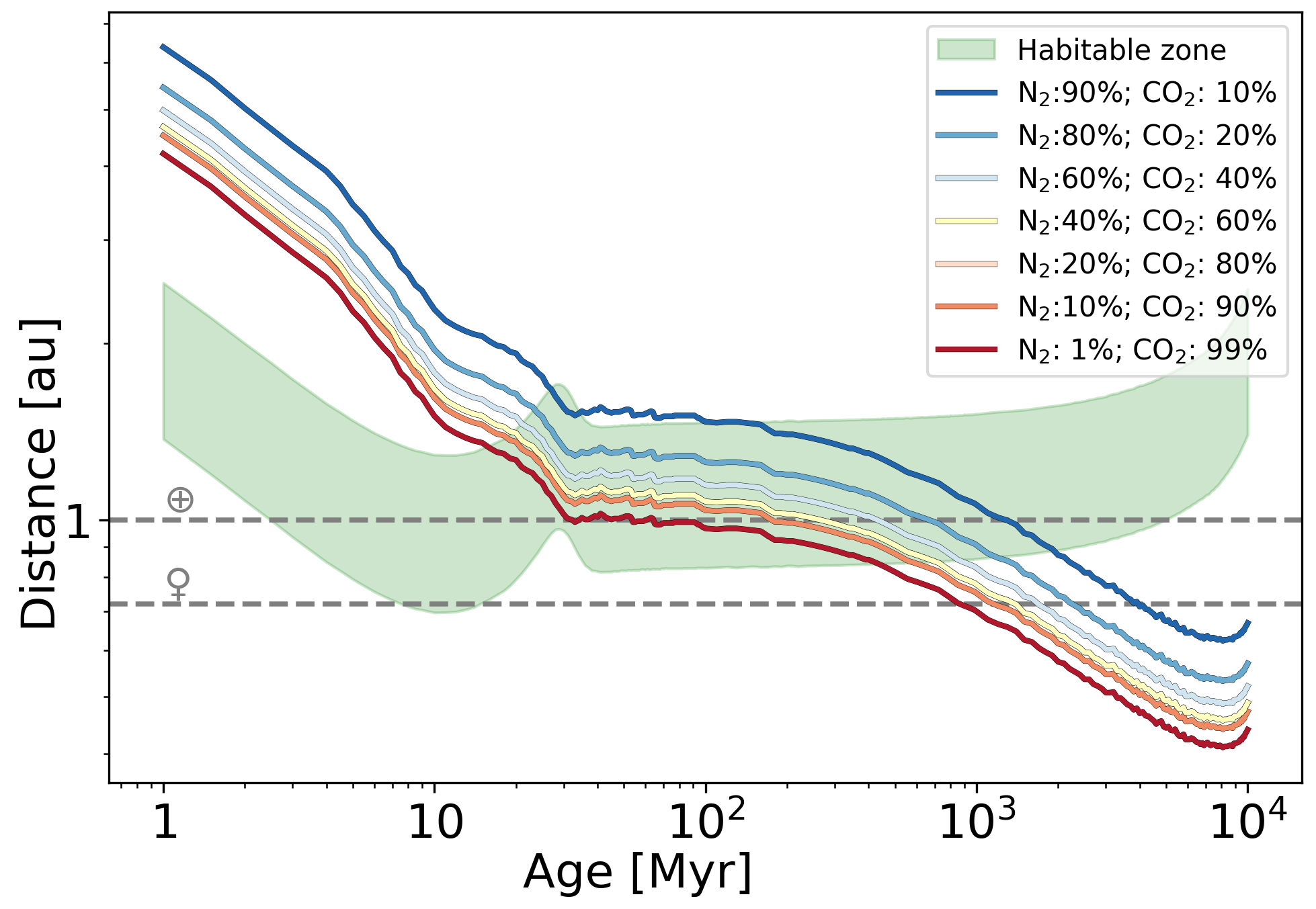}
        \caption{Atmosphere retention distance for different atmospheric compositions, indicated by different colours, over time for a 1 M$_\odot$ slowly rotating star. Above these lines, atmospheres can be retained. The green shaded area indicates the HZ. The dashed lines indicate Earth's and Venus's semi-major axis, denoted by their corresponding symbols.}
        \label{fig:dist_age_1p0}
    \end{figure}

In Fig. \ref{fig:dist_age_1p0} we show how the HZ, indicated by the green shaded area, changes over time for a slow rotator star of 1.0 M$_\odot$. In this figure, we also indicated the ARD of various compositions, marked by the different coloured lines. At smaller distances (i.e. below the coloured lines) the escape rate is greater, making the presence of an atmosphere even less likely. On the other hand, at distances greater than the retention distance an atmosphere can be maintained over long timescales or even increased if the outgassing rate is high enough.

Here we can see how the bolometric luminosity of a Sun-like star, which defines the HZ, does not evolve the same way as the EUV luminosity, which we used to determine the ARDs. While the bolometric luminosity stays almost constant over much of the star's main-sequence life and gradually increases towards the end of it, the EUV luminosity declines substantially over time due to stellar spin-down and the consequent weakening of the magnetic dynamo. Therefore, certain orbits that are habitable at one point might not be at a later point in time, though this zone does not change drastically over the lifetime of the Solar System. 

If we focus on the first 25 Myr in Fig. \ref{fig:dist_age_1p0}, we see that a 1 au orbit is closer than the ARD. It is unlikely that the secondary atmosphere is fully formed yet at this age. Over this short period, the integrated mass loss would add up to several tens of bars of atmosphere, which is likely less than the available atmospheric inventory. This high loss rate might be interesting when looking at the stripping of a primordial atmosphere, though the models presented in this work are not an adequate representation of hydrogen-dominated atmospheres. For studies on the mass-loss rates of hydrogen-dominated atmosphere, we refer the readers to works such as \citet{kubyshkina2018a} and \citet{owen2016}.

The period immediately following this highly irradiated time would coincide with the estimated Moon-forming impact between 30 and $\sim$100\,Myr \citep[e.g.][]{hopkins2008,yu2011,fu2023}. As summarised in \citet[and references therein]{gebauer2020} the Earth would have had a magma ocean following the impact, which they estimate would have led to the catastrophic outgassing of 70 bars of CO$_2$. In Fig. \ref{fig:dist_age_1p0} we can see that at this time, the mass-loss rate would still be of the order of 1 M$_{\mathrm{atm},\oplus}$/10 Myr, although it would decrease throughout the rest of the Hadean. Upon integrating the mass loss at Earth's distance from 100 Myr to 600 Myr, we find a total mass loss due to Jeans escape of the order of $3 \times 10^{19}$~kg (around 5.5 M$_{\mathrm{atm},\oplus}$). This is well below the initial atmospheric mass, in particular, if the Earth continued to release more volatiles as discussed in Sect. \ref{sec:outgassing}.

This loss rate would easily leave the Earth with over 60 bar of CO$_2$ at the start of the Archean. Isotope measurements of quartz from the early-Archean ($\sim$600-1600 Myr), point towards surface pressures of the order of just 1-2 bar \citep{marty2013}. Similarly from gas bubbles in basalt, the Earth is estimated to have had a partial CO$_2$ pressure of 0.23-0.5 bar by the late Archean ($\sim$1900 Myr) \citep{som2016}. It is important to re-emphasise that the mass-loss rates predicted by our model are only based on Jeans escape. The additional atmospheric mass compared to the geological record could have been lost to space through hydrodynamic escape \citep{tian2008a}, or non-thermal loss processes \citep{kislyakova2020,grasser2023}. Additionally, water and CO$_2$ would have been returned to the mantle through chemical weathering, dissolution into (magma) oceans, subduction, and sequestration into the crust \citep[as summarised in][]{lammer2018}.

\cite{lammer2018,lammer2019} summarise the complexity of the carbon and nitrogen cycles on Earth, which are strongly tied to the oceans, plate tectonics, and life. Estimates of atmospheric N$_2$ contents during the middle Archean (1100--1600 Myr) from fluid inclusions in quartz indicate a partial pressure between 0.5 and 1.1 bar \citep{marty2013}. The partial CO$_2$ pressure is estimated to be lower than 0.7 bar, meaning the atmosphere reached a roughly 50\%/50\%~CO$_2$/N$_2$ composition \citep{som2016}. Although there are large uncertainties associated with indirect measurements of the early atmosphere, it is interesting to note that this measurement is consistent with the models presented in Fig. \ref{fig:dist_age_1p0} as the ARD of the 60\% CO$_2$ atmosphere crosses 1au at 200 Myr. 

Venus is the other rocky planet with a thick atmosphere that is of interest for this discussion. At a distance of 0.73 au, Venus is closer in than the ARD up until 1000 Myr. Due to the very limited geological (or cytherological, if you will) data of Venus, we cannot constrain the early history of Venus as we can for Earth. \citet{Kreslavsky2015} analysed radar data of Venus and concluded that the planet underwent extended resurfacing at ages between 3600 and 4400 Myr (i.e. 1 and 0.2 Gyr ago). In that work, the authors also conclude that due to the size and distribution of craters, the atmospheric shielding must have remained similar throughout this period of Venus's history. Our models predict that the mass-loss rate during this period would be relatively low for a CO$_2$-dominated atmosphere. \citet{avice2022} summarise other methods that have been used to constrain the earlier history of Venus, including isotopic ratios. In particular, Venus's elevated isotopic ratio of D/H seems to indicate that it experienced greater atmospheric loss \citep{donahue1982}. If we assume the largest mass-loss rate over the period from 100 to 1000 Myr, the loss would amount to the order of $6 \times 10^{20}$~kg (around 124 M$_{{\rm atm},\oplus}$). However, this would be a lower limit as the irradiance at 0.72 au at 100 Myr is 24 F$_{{\rm EUV,}\oplus}$ and decreases to 13 F$_{{\rm EUV,}\oplus}$, which is greater than the maximum modelled irradiance. This loss of atmosphere could be avoided if the secondary atmosphere only built up gradually over an extended amount of time (beyond 1000 Myr) rather than in a catastrophic outgassing event. \citet{orourke2015} used isotope ratios of argon to demonstrate that a stagnant-lid model, a model in which the atmosphere is outgassed over several billion years, can explain the current environment of Venus.

These results demonstrate that there is a complicated interplay between the atmosphere and interiors of rocky planets. Adding to this, other intrinsically random events such as giant impacts leave us with great uncertainty when it comes to determining the early history of planets and the formation of secondary atmospheres. For Earth-like planets and orbital distances, global solidification is estimated to finalise between the ages of 0.1 to 100 Myr, although this may take longer for planets with a higher irradiance \citep[and references therein]{lichtenberg2023}. To make a more conservative estimate we suggest disregarding the first 1000 Myr to determine reliable ARDs. Through this assumption we want to encapsulate the possibility of early very dense atmospheres that erode to thinner atmospheres in early stages (as we surmise for Earth), and significant outgassing over prolonged periods of time (as might be the case for Venus). Of course, we could think of alternative scenarios that would allow the existence of atmospheres around planets that orbit their host star closer than the ARD (e.g. planetary migration, giant impact later than 1000 Myr), but to generalise our results we assume that it is unlikely that these planets would retain a significant atmosphere.

\subsection{Stellar mass} \label{sec:res_massvar}
\begin{figure*}
    \includegraphics[width=1.0\textwidth]{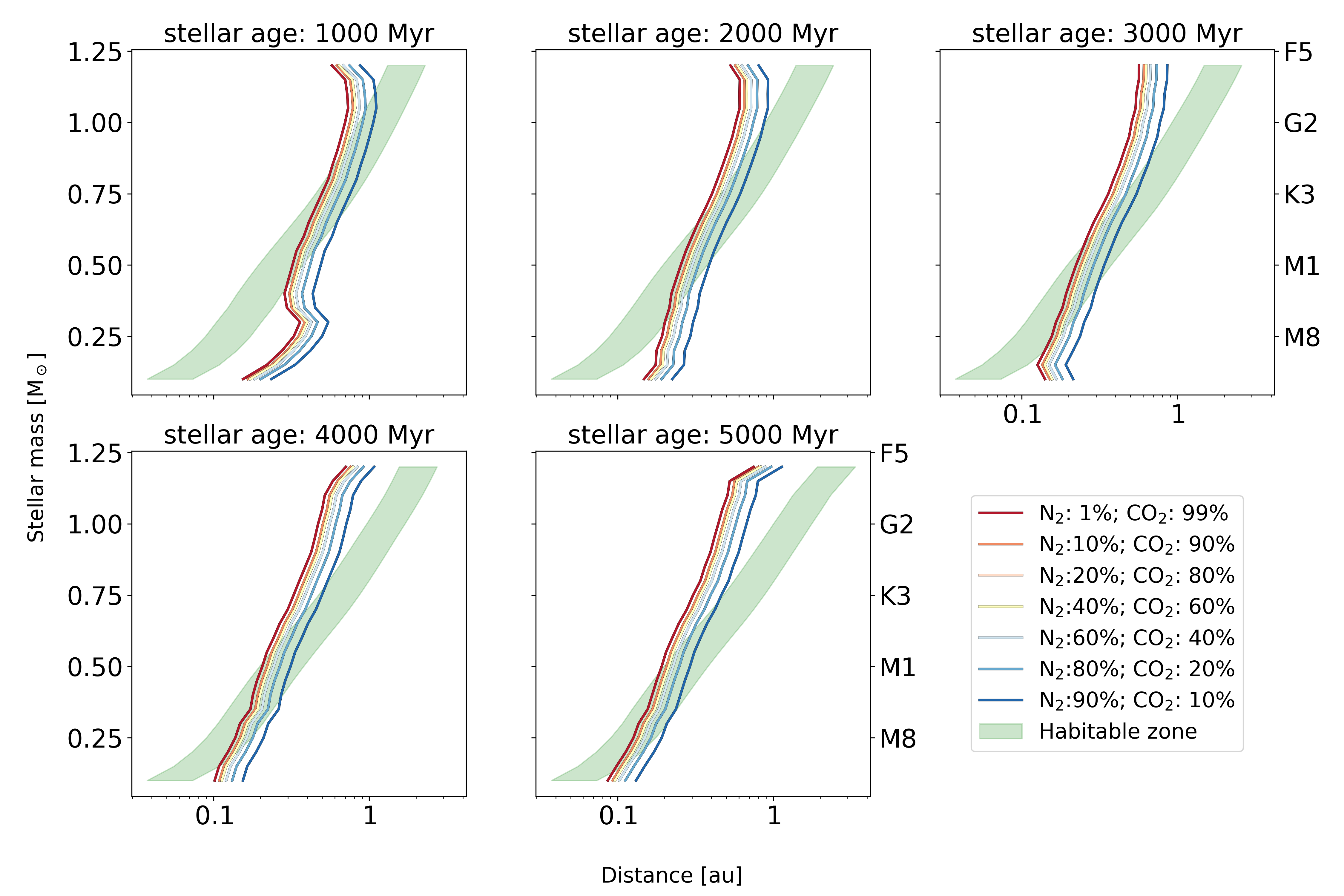}
        \caption{Atmosphere retention distance for different atmospheric compositions, indicated by different colours, for different stellar masses for slowly rotating stars. To the right of these lines, atmospheres can be retained. The green shaded area indicates the HZ. Each panel shows a different stellar age.}
        \label{fig:mass_dist}
    \end{figure*}

The different panels of Fig. \ref{fig:mass_dist} indicate different stellar ages. Each panel shows the HZ as a green-shaded area for various stellar masses, assuming an initially slowly rotating star. The various coloured lines indicate the ARD for different atmospheric compositions, similar to Fig. \ref{fig:dist_age_1p0}. Contrary to the time evolution plot, the distances where an atmosphere can be retained are to the right of the line and orbits where atmospheres would be lost are to the left of the line. When we look at stars beyond our Sun we can see that the HZ does not change in the same way as the ARD when we vary the mass of the star. In particular, for the lowest-mass stars, we can see that most atmospheric compositions cannot be retained within the HZ. Just as we saw in Fig. \ref{fig:dist_age_1p0}, younger stars are much more active in the EUV regime leading to ARDs that are beyond the respective HZs.

Figure \ref{fig:mass_dist} shows that at an age of 1000 Myr, the ARD of CO$_2$-dominated atmospheres and outer HZ only cross for stars with a mass greater than 0.4 M$_\odot$. For N$_2$-dominated atmospheres this only happens for stars with a mass greater than 0.6 M$_\odot$. Based on our knowledge of the inner Solar System planets and volcanism, we expect that the atmospheres of young rocky planets in and near the HZ are CO$_2$-dominated. When we examine the history of Earth's atmosphere, we can also see that these CO$_2$-dominated atmospheres can still evolve to N$_2$-dominated atmospheres if the surface conditions and tectonics allow for it. Due to this possible evolution, we first focused on the CO$_2$-dominated atmospheres.

At this early epoch, we can see that the EUV luminosity (and by extension the ARD) of stars with a mass smaller than 0.35 M$_\odot$, does not follow the same trend as that of the higher-mass stars. As discussed by \citet{johnstone2021a}, this difference is due to the much lower spin-down rate of fully convective stars. This results in an ARD at a higher orbital distance than the outer edge of the HZ. As demonstrated in the previous section, the mass loss due to Jeans escape can amount to tens and even hundreds of M$_{{\rm atm,}\oplus}$ for an orbit at the ARD at 1000 Myr. Venus shows us that either such losses can be recovered shortly after or the initial atmospheres can be large enough to offset such a loss. The exponential increase in the mass-loss rate shown in Fig. \ref{fig:mass_losses} illustrates that orbits closer than this distance would swiftly accumulate losses surpassing the total volatile content of an Earth-mass planet. This fact is reinforced further by the fact that the irradiance scales by a$^{-2}$, where a is the semi-major axis. This leads us to conclude that planets on orbits within or just outside of the HZ are unlikely to retain their atmospheres at this age.

When the planetary systems reach an age of 2000 Myr (top right panel of Fig. \ref{fig:mass_dist}), we can see that for stars with masses above 0.6 M$_\odot$ the entire HZ is further out than the ARD. On the low-mass end, we can see that the ARD is still further out than the HZ for stellar masses lower than 0.3 M$_\odot$. This means that planets in the HZ of these stars would have experienced mass-loss rates orders of magnitude greater than the outgassing rate discussed in Sect. \ref{sec:outgassing} for 2000 Myr. This further strengthens our conclusion that these planets are unlikely to retain any atmosphere.

The second and third rows of Fig. \ref{fig:mass_dist} show a trend of a slowly decreasing ARD whilst maintaining a relatively stable HZ. These panels at later ages can help limit the possible compositions of atmospheres. As an example we can take a closer look at the right panel of the middle row, depicting systems at an age of 4000 Myr. On the one hand, around a 1 M$_\odot$ star, any composition would have negligible losses. On the other hand, around a star of 0.5 M$_\odot$ a nitrogen-dominated atmosphere such as Earth's, would be stable near the outer edge of the HZ, but not near the inner edge, although this does not necessarily imply the presence of liquid water as mentioned in Sect. \ref{sec:hz}. On top of constraining the possible parameter space searched in observational programmes, knowing the possible bulk composition of planets can give us insights into habitability, particularly when we look at more complex life. \citet{catling2005} describe how a certain amount of free oxygen is likely required to provide sufficient energy for complex organic structures. As shown by \citet{alcott2024}, the buildup of free atmospheric O$_2$ linked to the capture of CO$_2$ back into the surface. Additionally, \citet{stueken2016} show that both high N$_2$ and O$_2$ levels together might be required for life. This makes the ARD of N$_2$-dominated atmospheres more important when considering the possibility of complex life.

We want to emphasise that our models could only be used to exclude certain combinations of stellar mass, orbital distance, and atmospheric compositions that would not be able to support long-term atmospheric retention. For example, we can see that around a 1 M$_\odot$ star at an age of 4600 Myr and an orbital distance of 0.72 au, a N$_2$-dominated atmosphere could be stable, but so could a CO$_2$-dominated atmosphere, as demonstrated by the current composition of the atmosphere of Venus.

\begin{figure}
    \includegraphics[width=\linewidth ]{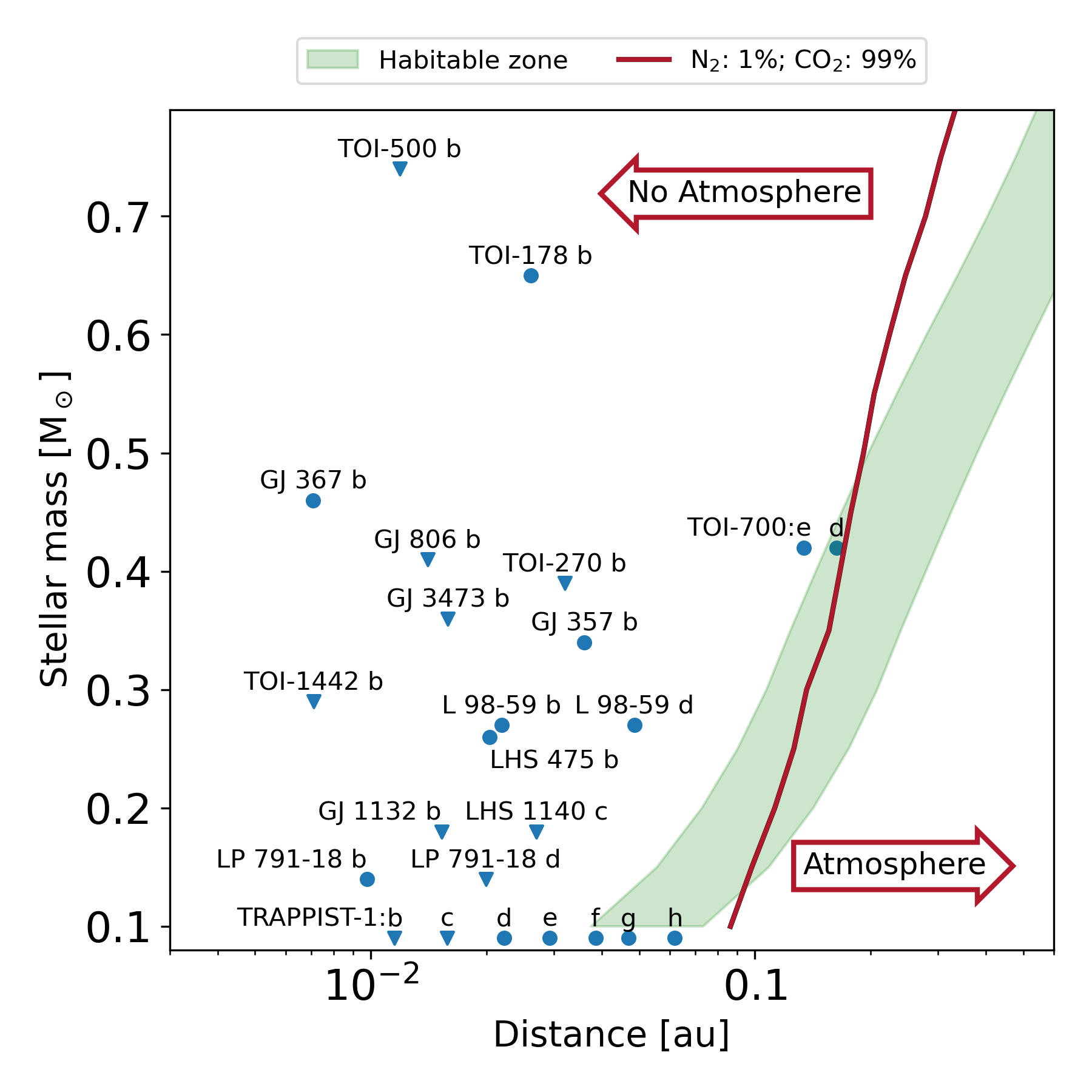}
        \caption{Atmosphere retention distances for different stellar masses for slowly rotating stars at an age of 5000 Myr. The green shaded area indicates the HZ. Symbols indicate scheduled JWST targets, with triangles highlighting eclipse observations. For clarity, the system's name is not repeated for each planet in the TRAPPIST-1 and TOI-700 systems.}
        \label{fig:observations}
    \end{figure}

Lastly, we compared our results to the observations currently scheduled for the JWST mission. In \citetalias{vanlooveren2024} we noted the dependence of the mass-loss rate on the mass of the planet when varying the masses between 0.8 and 1.2 M$_\oplus$. As such, we only selected the scheduled observations of planetary masses close to our models (i.e. 1 M$_\oplus$). \citet{rogers2015} shows that the boundary between rocky planets and planets with extended gaseous envelopes lies at 1.6 R$_\oplus$. To allow for some error in mass measurement, we selected all JWST targets with masses up to 2 M$_\oplus$. Figure \ref{fig:observations} is a repetition of the distance against stellar mass plot at 5000 Myr (bottom left panel of Fig. \ref{fig:mass_dist}) with these targets added. Eclipse observations are indicated with triangles as this observation method allows us to differentiate more clearly between planets with and without atmospheres. At the bottom of the figure, we see the TRAPPIST-1 planets, which are all closer to their host star than the ARD \citep{vanlooveren2024}. We can see that most of these Earth-sized planets around more massive stars orbit much closer than the ARD. This is to be expected due to the observational bias towards short-period planets. Figure \ref{fig:observations} does indicate that it is unlikely that any of these planets would retain any atmosphere for an extended period of time unless it is rapidly replenished through continued and excessive outgassing. A detection of atmospheres around these objects would then indeed suggest strong outgassing activity, for example through volcanism. 

For stars at the low end of our mass range, the actual atmospheric loss rates are likely even greater than this figure indicates. Due to the relatively low luminosity of these stars given their small surface area and effective temperature compared to, for example, the Sun, their flares can reach luminosities similar to or greater than the bolometric luminosity of the star \cite[e.g.][]{ealy2024}. This sudden increase in flux, in particular in XUV, can change the chemical composition of the atmosphere leading to an increased thermal loss rate \citep[e.g.][]{segura2010,doamaral2022, chen2021}. Though the short duration of flares leaves the long-term influence of large flares uncertain \citep[e.g.][]{nicholls2023}, it is important to keep in might that factors not taken into account in this work might further influence these mass-loss rates.

The target closest to the ARD is TOI-700 d, a 1.72 M$_\oplus$ planet orbiting within the HZ of its host star \citep{suissa2020}. On the one hand, as the mass of the planet is higher than our modelled mass, the actual ARD could be closer. On the other hand, because the star could be as young as 1.5 Gyr the ARD could be much further out and closer to the ARD shown in the first panel on the right column of Fig. \ref{fig:mass_dist}. In a previous study of this target, \citet{nishioka2023} found that the ion escape, a type of non-thermal loss, would erode a Venus-like atmosphere in the absence of a strong intrinsic magnetic field. Although only the observations currently scheduled for 2025 will bring more certainty, models seem to indicate that this planet is unlikely to retain any atmosphere.

\subsection{Stellar evolution and rotation} \label{sec:res_evol}
\begin{figure*}
   \centering
    \includegraphics[width=1.0\textwidth ]{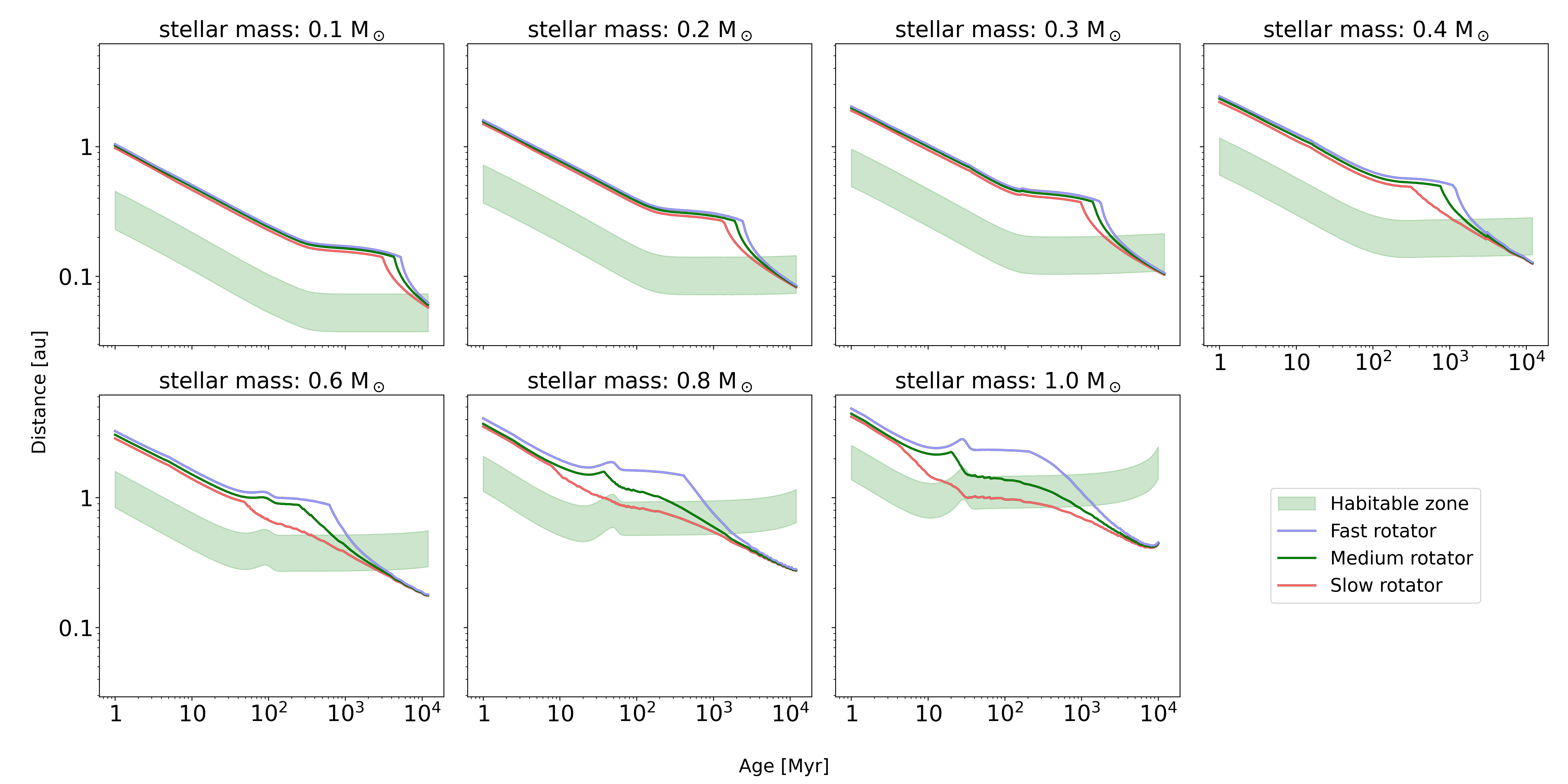}
        \caption{Atmosphere retention distances for slow rotators (red), medium rotators (green), and fast rotators (blue) for a 1 M$_\oplus$ planet with a 99\% CO$_2$ atmosphere. Above these lines, atmospheres can be retained. The green shaded area indicates the HZ. Each panel shows a different stellar mass.}
        \label{fig:dist_age}
    \end{figure*}

Each panel of Fig. \ref{fig:dist_age} shows a different stellar mass. Similar to Fig. \ref{fig:dist_age_1p0}, in each panel the green shaded area denotes the evolution of the HZ over time. In the models shown in Fig. \ref{fig:dist_age}, we assumed a planet of 1 M$_\oplus$ with an atmosphere composed of 99\%~CO$_2$ and 1\%~N$_2$. The different lines indicate the ARD for different initial rotation rates with the slow rotator in red, the medium rotator in green, and the fast rotator in blue. Some important points to notice are: (1) the HZ remains the same for all initial rotation rates as the bolometric luminosity is not affected in any significant way by the initial rotation rate; and (2) this figure shows a similar evolution to Fig. 11 from \citet{johnstone2021a} as the ARD is calculated based on the EUV regime, which scales with the X-ray luminosity. Figure \ref{fig:dist_age} also demonstrates that the initial rotation rate is not equally important for each stellar mass.

If we focus on the 1 M$_\odot$ case, we can see very different ARDs depending on the initial rotation rate. This difference was discussed in detail in \citet{johnstone2021} and using geological records of the Archean atmosphere, the authors concluded that the Sun must have formed as a slow to medium rotator. In the more generalising approach we used in this work, we could conclude from this figure that only the outer half of the HZ of a fast-rotating Sun-like star would be hospitable for atmospheres between 1000 and 1500 Myr. For an intermediately rapidly rotating star, the entire inner edge of the HZ crosses the ARD at 1000~Myr. Therefore, if the Sun was born as an intermediate rotator, the Earth could still retain an atmosphere but would have followed an entirely different evolutionary path than the one indicated by geological records.

As we shift our focus to lower-mass stars, we can see that the division between different initial rotation rates shifts towards later ages. In the panel for a 0.4 M$_\odot$ star, we can see that the ARD of the slowly rotating star overlaps with the outer edge of the HZ at 1000 Myr, meaning that a planet this distance could hold onto an atmosphere. For a similar star with a fast initial rotation rate, the ARD would only pass the outer edge of the HZ at 2000 Myr. This means that the same planet would experience a 1000 Myr of extreme mass loss, making it less likely to have retained an atmosphere. For the fully convective stars, this gap increases even further due to their slow spin-down rate. In the panel for a 0.1 M$_\odot$ star, the ARD crosses the outer edge of the HZ at 7000 Myr for an initially slowly rotating star and at 9000 Myr for an initially fast rotating star. Although in neither case the planets within the HZ of such a star are likely to have retained any atmosphere, this can become more important for planets on wider orbits.

All of these results highlight that, even though the stellar mass has a greater impact on the ARD and the likelihood that planets can maintain their atmospheres, the initial stellar rotation rate proves to add a layer of complexity to determining the likelihood of atmospheres' survival. This complexity is further increased by the fact that we can only observe the current rotation rate of a star, which does not necessarily allow us to determine the initial rotation rate and evolutionary path of the star, as is the case for our Sun \citep{tu2015}. This is because all evolutionary tracks of rotation converge to nearly the same track at some evolutionary stage, making backwards tracing difficult.

\section{Conclusions}\label{sec:conc}
In this work we determined the range of conditions under which an Earth-mass planet could retain its atmosphere in the HZ. To achieve this goal, we combined the mass-loss rates due to Jeans escape modelled in \citet{vanlooveren2024} with the stellar rotational evolution models from \citet{johnstone2021a}. 

The mass-loss rates for different atmospheric compositions were used to determine the EUV irradiances for which the Jeans escape would become significantly greater than estimated outgassing rates, which we set to be of the order of $10^{4}$~kg/s. This irradiance could then be converted into an ARD for different stars based on their EUV luminosities.

First, we validated our models against the Solar System. For Earth, we see the need for a CO$_2$-dominated atmosphere in its early history, which is also necessary to explain the faint young Sun paradox \citep{feulner2012}. This requirement becomes less stringent at 1000 Myr. This result is in good agreement with earlier work by \citet{johnstone2021}, who determined a minimum mixing ratio of 40\% CO$_2$ for the Earth during the early Archean ($\sim$600 Myr). Our models also show that Venus could have retained a dense atmosphere until the start of the last global resurfacing ($\sim$3500 Myr ago; \citealt{Kreslavsky2015}). Taking into consideration what we know of the Solar System planets from geological evidence,  we assume that phenomena such as global magma oceans, planetary migration, or initial atmosphere could offset the losses that happened during the first 1000 Myr. 

We then turned to other stellar masses by calculating the ARD based on the EUV luminosity of these stars at different ages. Due to their low spin-down rates, the ARD of fully convective stars (M < 0.35 M$_\odot$) does not overlap with the HZ after the 1000 Myr grace period for any of the modelled atmospheric compositions. This implies that Earth-mass planets in the HZs of these stars are unlikely to retain atmospheres unless they are subject to excessive outgassing. Looking at later ages, these models can help constrain possible bulk compositions of atmospheres, similar to the Archean Earth.

Additionally, we looked at the influence of the initial rotation rate of the star on the ARD. We conclude that the initial rotation rate can influence the ARD and thus the probability of a planet maintaining an atmosphere, in particular closer to the inner edge of the HZ. This additional parameter adds a layer of complexity, in particular for stars at ages when the different rotational tracks merged again, making it impossible to assess the early environment of the stellar system.

Finally, we compared our model results against JWST cycle 1 and 2 Earth-sized rocky exoplanets and show that none of these targets are expected to retain an atmosphere. More generally, we conclude that HZ Earth-mass planets are most likely to retain their atmospheres around (preferably slowly rotating) stars with masses greater than 0.4 M$_\odot$.

\begin{acknowledgements}
      The results of this work were partially achieved at the Vienna Scientific Cluster (VSC).
      We acknowledge the Community Coordinated Modeling Center (CCMC) at Goddard Space Flight Center for the use of the Instant Run System of the NRLMSIS model, https://kauai.ccmc.gsfc.nasa.gov/instantrun/nrlmsis/.
      The project was partially funded by the European Union (ERC, EASE, 101123041). Views and opinions expressed are however those of the author(s) only and do not necessarily reflect those of the European Union or the European Research Council Executive Agency. Neither the European Union nor the granting authority can be held responsible for them.
\end{acknowledgements}

%
%
\bibliographystyle{aa}
\bibliography{citations.bib}

\begin{appendix} 

\end{appendix}
\end{document}